\PassOptionsToPackage{pdfpagelabels=false}{hyperref}

\documentclass{sigchi}

\toappear{
}

\usepackage{balance}  
\usepackage{graphics} 
\usepackage{txfonts}
\usepackage{times}    
\usepackage{color}
\usepackage[usenames,dvipsnames]{xcolor}
\usepackage{textcomp}
\usepackage{booktabs}
\usepackage{ccicons}
\usepackage{ifthen}
\usepackage{enumitem}
\usepackage{graphicx,dblfloatfix}
\usepackage{xspace}

\newcommand{\hide}[1]{}

\iffalse

\newcommand{\peter}[1]{\textcolor{PineGreen}{[Peter: #1]}}
\newcommand{\shang}[1]{\textcolor{blue}{[Shang: #1]}}

\newcommand{\moushumi}[1]{\textcolor{blue}{[Moushumi: #1]}}

\else

\newcommand{\peter}[1]{#1}
\newcommand{\shang}[1]{#1}

\newcommand{\moushumi}[1]{#1}

\fi

\newcommand{\focal}[0]{\textit{focal}\xspace}

\newcommand{\AND}[0]{\textbf{AND}\xspace}
\newcommand{\Orchestra}[0]{\textit{Event Orchestra}\xspace}
\newcommand{\Stage}[0]{\textit{Sequence Stage}\xspace}
\newcommand{\SEQone}[0]{$\vcenter{\hbox{\includegraphics[width=0.45in]{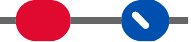}}}$}
\newcommand{\SEQtwo}[0]{$\vcenter{\hbox{\includegraphics[width=0.2in]{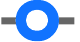}}}$}

\newcommand{\SEQfour}[0]{$\vcenter{\hbox{\includegraphics[width=0.19in]{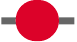}}}$}
\newcommand{\SEQfive}[0]{$\vcenter{\hbox{\includegraphics[width=0.6in]{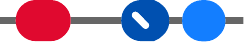}}}$}
\newcommand{\SEQsix}[0]{$\vcenter{\hbox{\includegraphics[width=0.2in]{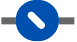}}}$}
\newcommand{\SEQseven}[0]{$\vcenter{\hbox{\includegraphics[width=0.2in]{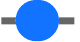}}}$}

\makeatletter
\def\url@leostyle{%
  \@ifundefined{selectfont}{\def\UrlFont{\sf}}{\def\UrlFont{\small\bf\ttfamily}}}
\makeatother

\def\pprw{8.5in}
\def\pprh{11in}

\definecolor{linkColor}{RGB}{6,125,233}

\begin{document}

\title{Chronodes: Interactive Multi-focus\\ Exploration
of Event Sequences}

\numberofauthors{2}
\author{%
  Peter J Polack Jr, Shang-Tse Chen, Minsuk Kahng,\\Kaya de Barbaro, Rahul Basole, Duen Horng Chau\\
  College of Computing\\
  Georgia Institute of Technology\\
  \email{ \{ppolack, schen351, kahng, kaya, basole, polo\}@gatech.edu}
    \and
    Moushumi Sharmin\\
    College of Science \& Engineering\\
    Western Washington University\\
    \email{moushumi.sharmin@wwu.edu}
}


\maketitle

\begin{abstract}
The advent of mobile health (mHealth) technologies challenges the capabilities of current visualizations, interactive tools, and algorithms.
We  present  Chronodes, an interactive system that unifies data  mining  and human-centric visualization techniques to support explorative analysis of 
longitudinal mHealth data.
Chronodes extracts and visualizes frequent event sequences that reveal
chronological patterns across multiple participant timelines of mHealth data.
It then combines novel interaction and visualization techniques to enable multi-focus event sequence analysis, which allows health researchers to interactively define, explore, and compare groups of participant behaviors using
event sequence combinations.
Through summarizing insights gained from a pilot study with 20 behavioral and biomedical health experts, 
we discuss Chronodes's efficacy and potential impact in the mHealth domain.
Ultimately we outline important open challenges in mHealth, and offer recommendations and design guidelines for future research.
For a video demonstration of Chronodes, please refer to the provided video figure.

\end{abstract}


\category{H.5.m.}{Information Interfaces and Presentation
(e.g. HCI)}{Miscellaneous}

\section{Introduction}
\begin{figure}[!t]
\centering
  \includegraphics[width=\columnwidth,]{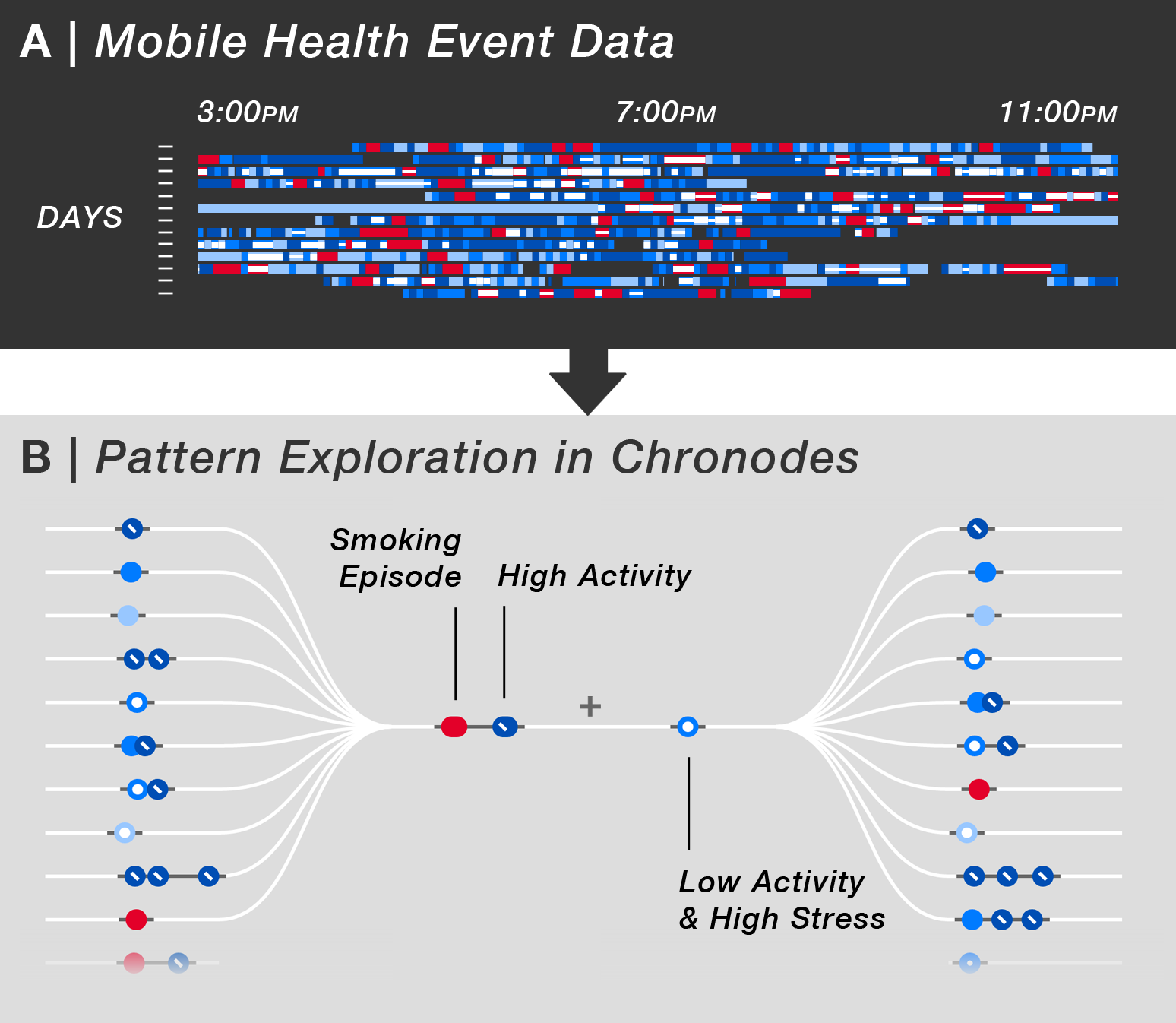}
  \caption{
  From mobile health (mHealth) event data, Chronodes extracts frequent patterns of events that users can interactively explore.
  (A) Multiple days of mHealth sensor data are represented as discrete events over time. Only a snapshot of the full dataset is shown.
  (B) Chronodes mines frequent sequences of events that  users can interactively combine to reveal patterns that occur before and after them (left and right).
  All figures best viewed in color.
  Red: smoking episodes. Blue: activity intensity.
  }~\label{fig:figure1}
\end{figure}

As revolutionary developments in mobile sensor technology extend to more applications in healthcare, the potential for understanding patterns and relationships among physiological factors is burgeoning.
With these sensors, health researchers are afforded new capabilities to collect and analyze health data from study participants, with more accuracy and precision than ever before.
We have commenced a pioneering initiative to develop innovative methods that turn the wealth of mobile health (mHealth) sensor data unlocked by these emerging and evolving sensors into reliable and actionable health information.
\moushumi{As smoking is the leading yet preventable cause of death in the United States, and as smoking lapse contributes to high failure rate in quit attempts, smoking cessation is one of the most challenging research problems in this area.}
\peter{For this reason, we have selected smoking cessation and lapse as our primary topic of focused analysis, and}
are currently developing tools to assist health researchers discover \moushumi{temporal,} physiological, and behavioral patterns and factors that cause abstinent smokers to lapse.

To assess the behaviors surrounding smoking abstinence and relapse, health researchers \moushumi{need to} to understand lifestyle choices \moushumi{and activities} \moushumi{that are related to} smoking relapse, and are particularly interested in the sequences of events leading up to lapse \cite{brandon1990postcessation,niaura1988relevance}. 
Furthermore, whereas analysts are interested in determining universal indicators and causes of smoking relapse, behavioral variation between participant groups is equally relevant \cite{cohen1990perceived, al2005attenuated}.
Should younger individuals' daily routines be distinguished from those of older?
Will early morning and late night smokers exhibit different pre-smoking behaviors \cite{patton1998course}?

\moushumi{To begin answering these questions, }
researchers need tools to discover which subsets (cohorts) of abstinent smokers exhibit similar or dissimilar behaviors, and to compare the groups as such.
Ultimately, these methods must also represent longitudinal, high-resolution mHealth datasets.
Figure \ref{fig:figure1} shows a snapshot of this complexity for only a few evenings' worth of mHealth data---\peter{visualizing the data as-is is complex and difficult to make sense of.}
Exploring how chronological patterns relate to common behaviors is non-trivial, as it is not only a challenge to define groups of behaviors from mHealth data, but also to examine and compare these groups interactively.
\peter{In contrast to traditional electronic health records (EHR) where temporal events are already explicitly defined, largely non-overlapping, and do not repeat frequently throughout the day, mHealth analysis requires that event chronology is derived from overlapping sensor data streams that span daytime continuously.}

\subsubsection{Mobile Health (mHealth) Datasets}

We obtained real-world mHealth datasets collected through \moushumi{field studies} 
investigating the use of the AutoSense Sensor Suite 
for inferring general stress state \cite{ertin2011autosense}.
In the first of these studies, 6 participants (who smoke) wore the sensors for 3 days, totaling about 400 hours.
\peter{In a proceeding study, the same format of data was recorded for 52 participants before and after planned smoking abstinence (pre-quit and post-quit, respectively), with each participant's data totaling approximately 6 days} \cite{saleheen2015puffmarker}.
The sensor suite recorded 40 data streams for each participant, capturing a wide variety of physiological signals, 
e.g., electrocardiogram (ECG), galvanic skin response (GSR), and heart-rate variability (HRV) measurements. 
The resulting datasets consisted of over 750,000 and 4,650,000 data points, respectively (sampled at 1 Hz).
The AutoSense data represents one of the few mHealth data initiatives available for research use, as collecting good-quality mHealth data is a great technical challenge, due to limitations in sensor hardware and battery technologies \cite{ertin2011autosense, cios2002uniqueness, grimson2000si}.

\subsubsection{Our Contributions}


\begin{itemize}[labelindent=1cm, itemsep=0cm,
topsep=0cm]





\item 
MHealth data introduces unique challenges that existing tools for analyzing EHR do not adequately address. 
Chronodes is one of the first attempts to explore how to best represent and explore mHealth data, combining data mining and human-centric visualization techniques to aid 
data exploration, 
pattern discovery,
and fine-grained analysis of longitudinal mHealth records.
\item
Chronodes introduces two novel capabilities that help  health researchers interactively explore and combine sequences to ask important, complex questions like \textit{``what happens between first lapses that precede high activity, and second lapses that follow low activity but high stress?''}, and then to generate and explore variations of this question interactively:
\begin{enumerate}
\item While existing tools align on multiple occurrences of a single event type, Chronodes allows alignment on two or more event types so users can see what happens in between (e.g., between first and second lapse). 
\item Chronodes' alignment points can be a single event, or a sequence consisting of multiple events (e.g., smoking lapse $\rightarrow$ high-activity). 
\end{enumerate}


\item We conducted an informal study with 20 expert researchers from a spectrum of mHealth related disciplines. \moushumi{By }summarizing the insights gained, we outline important open research and design challenges in 
developing the mHealth field, and offer recommendations and design guidelines for future research.
\end{itemize}
\hfill\\
For a video demonstration of Chronodes, please refer to the provided video figure.



\section{Research Questions \& Design Motivations}

Before beginning development on Chronodes, we worked with health researchers on our  team to collect important questions that an interactive visualization tool should help answer:
\\

\begin{figure*}[!t]
\centering
  \includegraphics[width=\textwidth]{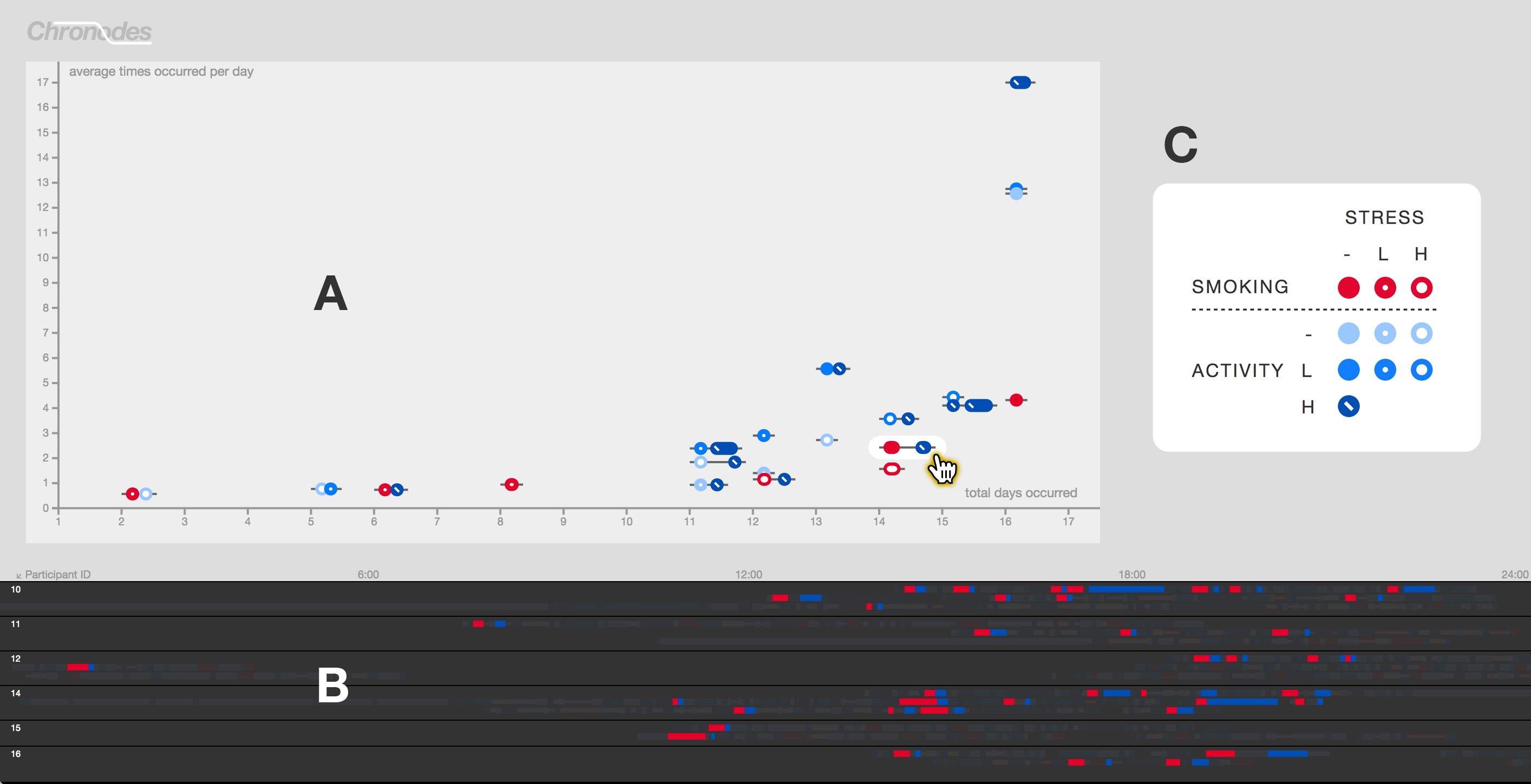}
  \caption{(A) The Chronodes \Stage showing a scatterplot of frequent sequences, and (B) the \Orchestra below it displaying the participant event data that these sequences were mined from.
  Whereas all sequences in the \Orchestra are grayed out by default, when the user mouses over a frequent sequence on the \Stage its location throughout the participant data is highlighted in the \Orchestra.
  (C) The legend for this pilot dataset is magnified here for better readability.
  }~\label{fig:fullscreen}
\end{figure*}

\begin{description}[labelindent=0cm, leftmargin=0.7cm, itemsep=0cm, topsep=-0.15cm]

\item[Q1] What are the events preceding and proceeding each instance of smoking lapse? \cite{brandon1990postcessation}

\item[Q2] What habitual events or cues (e.g., smoking every day after lunch) are correlated to smoking lapse? \cite{niaura1988relevance}

\item[Q3] What are the correlations between smoking lapse and other physiological factors, such as stress? \cite{cohen1990perceived, al2005attenuated}

\item[Q4] What event patterns are specific to individual participants, and otherwise universal to all participants? \cite{patton1998course}
\end{description}
\hfill\\
We propose to leverage \textit{frequent sequence mining} as a core method to approach these research questions.
Mining mHealth data for recurring event sequences helps to answer questions Q2 and Q4: it permits investigations of participant behaviors within days, across days, and between participants, all at a generalized, human-interpretable level.
By then adapting `\textit{event} alignment' into `\textit{frequent sequence} alignment,' we enable answers to Q1: we allow Chronodes users to investigate the frequent sequences that occur before and after aligned sequences of interest.

To address Q3, finding correlations between smoking events and physiological factors, we can represent physiological factors as events.
For example, we can consider a high-stress activity and a drive to work as different events, and represent them within frequent sequences once they are mined.
As we intend for Chronodes to be applied to a wide variety of health conditions, it is important that the process of converting physiological factors to events is user-adjustable.
In this way, we enable users to adjust event definitions interactively as in \cite{fails2006visual}, before the events are mined for \peter{event sequence} patterns.
How these events are derived, and the implications of defining events in this way, are discussed in later sections.

\section{Chronodes Overview}
To preface Chronodes' contributions, we provide a high-level overview of the system's fundamental features.

\subsection{Description of the User Interface}

The Chronodes user interface (Figure \ref{fig:fullscreen}) is comprised of two areas that update to the user's interactions.
The \Orchestra, fixed to the bottom of the screen (Figure \ref{fig:fullscreen}B), lists \peter{a subset of} participants represented by our dataset, and displays the series of events that these participants perform over the course of 24 hour days.
As the user interacts with event sequences, participants and events that are associated with each sequence `light up' correspondingly (mousing over \SEQone\:in Figure \ref{fig:fullscreen}).
The \Orchestra demonstrates to new users how derived event sequences relate to the `raw' event data, and also serves as a continuous indicator of the subset of data selected elsewhere in the interface (highlighted).
\peter{In this way, the \Orchestra's primary role is as a guiding frame of reference, but is not essential to interacting with Chronodes and can be minimized.}

The \Stage is the primary area of user interaction, and accounts for the largest amount of space in the Chronodes interface (Figure \ref{fig:fullscreen}A).
It is within the \Stage that the user selects, manipulates, and compares frequent event sequences.
To demonstrate how the user performs these operations, we will describe Chronodes functionality through a use case scenario.
The provided video figure offers an overview of this scenario as well.

\subsection{Use Case Walk-through}
This scenario demonstrates how the pilot study data described in the Introduction is used with Chronodes in practice.
Our user Jane is a health researcher that intends to discover and understand physiological events commonly associated with smoking episodes.
Upon launch, Chronodes displays the events that occur throughout the participants' days in the \Orchestra.
In the \Stage, Chronodes displays the frequent event sequences derived from this event data.
By \peter{initially} displaying both the raw event data and the derived sequences side-by-side as in \cite{vrotsou2009activitree}, we lend Jane \peter{introductory} information about how event sequences are derived, what they represent, and where they occur.



\begin{figure}[!t]
\centering
  \includegraphics[width=\columnwidth,]{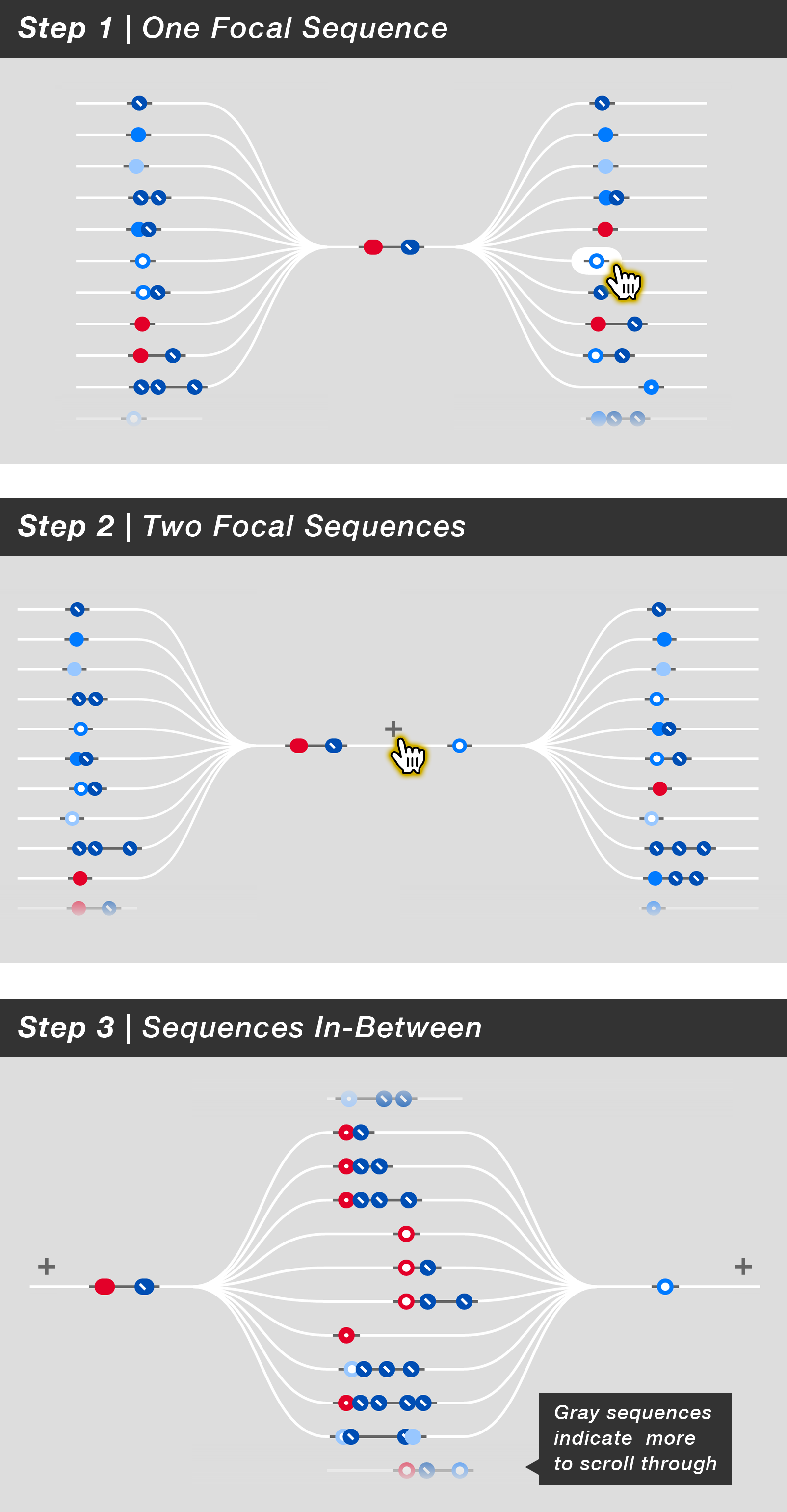}
  \caption{
  Varying configurations of \textit{focal} sequences and the sequences of events that occur around them. (Step 1) one \textit{focal} sequence with the sequences before and after it. The user clicks a sequence after and it becomes a second \textit{focal} sequence in (Step 2) the two \textit{focal} sequences with the sequences before and after them. The user clicks the \textbf{+} symbol and the timeline becomes (Step 3): the two \textit{focal} sequences with the sequences between them.
  }~\label{fig:steps}
\end{figure}

Initially, the frequent sequences are displayed in a scatterplot (Figure \ref{fig:fullscreen}A) that helps Jane to identify frequent sequences of interest.
The axes of this scatterplot will be described in detail later on; for now, Jane is interested foremost in events surrounding smoking episodes, so she chooses a smoking $\rightarrow$ high-activity event sequence (\SEQone\:in Figure \ref{fig:fullscreen}A) from the scatterplot.
As she moves the cursor over this sequence, the \Orchestra highlights the multiple occurrences of this sequence across multiple participants and days of data.
By then clicking the sequence, it becomes a \textit{focal} sequence, and moves to the center of the \Stage as all other sequences fade away.
Once centered, Chronodes reveals the \textit{adjacent} event sequences that occur most frequently before and after the smoking event sequence that Jane selected (Figure \ref{fig:steps}: Step 1).

To indicate which of these sequences occurs before and after the \textit{focal} sequence, Chronodes displays the adjacent sequences as on a timeline:
sequences that occur before the \textit{focal} sequence are placed on the left, and those that occur after are on the right (Figure \ref{fig:steps}).
The adjacent sequences are then ranked vertically by frequency, so that the most frequent ones are topmost.

At this point, sequence frequency is not the only thing that matters to Jane; distance in time from the \textit{focal} sequence is also relevant. 
To account for this, Chronodes positions sequences horizontally based on their proximity in time to the \textit{focal} sequence.
In other words, sequences that tend to occur immediately before or after the \textit{focal} sequence are closer to it (notice the horizontal staggering between sequences in Figure \ref{fig:steps}: \shang{Step 1}). 

Critically, it is important for our users to understand this at a glance, that chronology on the \Stage timeline is \textit{relative} to the \textit{focal} sequence, and not absolute as in a normal timeline.
\peter{This is distinct from existing healthcare analysis visualizations, which depict participant timelines with linear temporality.}
To demonstrate this visually, Chronodes displays the frequent sequences on white \textit{tracks} that depict the flow of time (Figure \ref{fig:steps}).
In this way, every \textit{track} that Jane can trace with her eyes is a sequence of sequences that occurs somewhere in the participants' data.
The \textit{tracks} `bottleneck' at the \textit{focal} sequence because it is the \textit{focal} sequence that every other sequence has in common in its proximity.

To reduce visual clutter, only the top 10 most frequent sequences on each side are displayed, but Jane can scroll vertically through these options to reveal more.
Jane sees a low-activity high-stress event (\SEQtwo\:in Figure \ref{fig:steps}: Step 1) that occurs only sixth most frequently, but immediately, after her \textit{focal} sequence.
She selects it, and it becomes a second \textit{focal} sequence along the \textit{tracks} (Figure \ref{fig:steps}: Step 2). 

Now that Jane has selected two \textit{focal} sequences, Chronodes displays the sequences that occur only where both \textit{focal} sequences are present (i.e. \SEQone$\:\cdots\:$\SEQtwo).
To illustrate this comprehension to \peter{new} users, the events satisfied by Jane's two \textit{focal} sequences remain dimly lit in the \Orchestra (as in Figure \ref{fig:fullscreen}: Step 2).
Although the adjacent sequences before and after these \textit{focal} sequences are still visible, another feature now emerges: Jane can click a \textbf{+} symbol in between the \textit{focal} sequences (Figure \ref{fig:steps}: Step 2) to reveal the sequences that occur \textit{between} two \textit{focal} sequences (Figure \ref{fig:steps}: Step 3).
This functionality reveals a fundamental advantage of interactive frequent sequence alignment: the capacity to reveal the trends that occur between two or more events of interest.

Jane can clearly see the frequent sequences that occur before, between, and after the two \textit{focal} sequences, but which participants do these adjacent sequences belong to?
What time of the day are they occurring?
As before, Jane can simply mouse over any sequence to reveal its associated events in the \Orchestra.
By doing so, Jane discovers that the sequences she is observing are shared by a certain subset of participants.
Furthermore, for each of these participants, the events she is highlighting tend to occur in the evening.
This shows that by defining these two \textit{focal} sequences, Jane has effectively navigated to a `cohort' of participants that mutually exhibit the event behaviors she specified.
This demonstrates the interactive and expressive capabilities of Chronodes, to use event patterns to discover cohorts of participants.

Finally, Jane can delete one of her \textit{focal} sequences to return to a previous state of her analysis.
As adding more \textit{focal} sequences defines a more specific cohort because it resembles an \AND operation, deleting \textit{focal} sequences, conversely, broadens the cohort.
Jane can continue to explore subsets of her participants' behaviors in this fashion.

\section{Chronodes Contributions}
Now that an overview of the Chronodes interface has been presented, we proceed in this section to describe these features in terms of the contributions they lend to research in visualization, data mining, and mHealth analysis.

\subsection{Visualizing Frequent Sequences for Interaction}


Chronodes develops on event-based timelines by replacing events with sequences of events that can be rearranged interactively. Every event sequence, whether discovered by frequent sequence mining or defined interactively by the user, is an interactive element.
As in `event alignment,' any event sequence can be designated as a \textit{focal} sequence, so that all events before and after it are displayed chronologically (Figure \ref{fig:steps}).

Representing mined frequent sequences for interaction \peter{is a novel integration of \shang{data mining} and visualization} with a series of conceptual and computational challenges that Chronodes is designed to solve.
Fundamentally, it is important for our users to understand that, although we represent frequent sequences as singular visual elements, they in actuality represent many event sequences over time.
\peter{This relationship is demonstrated by the \Orchestra: when the user mouses over a frequent sequence on the \textit{Stage}, the corresponding event sequences across any number of participants are highlighted in the \textit{Orchestra}.}
In other words, a frequent sequence in the \Stage inherently represents many event sequences that frequently occur in the \Orchestra.


We represent frequent sequences as a `kebab' (\SEQone), so as to convey the fact that frequent sequences are a series of events on a timeline, \peter{potentially separated by time.}
Therefore, the length and distancing of events within the frequent sequences (\textit{intra}-sequence, within the \SEQone) is determined by averaging the timestamps of these events, wherever they occur in the associated event sequences.
On the other hand, as every frequent sequence represents event sequences that occur at variable times, indicating chronology between sequences (\textit{inter}-sequence) is not as simple as averaging the timestamps of the constituent sequences together.
Although an average of event sequence timestamps, unlike a median, preserves inter-sequence chronology, we need to ensure that the average does not oversimplify what our users need to see during alignment.
For instance, if event sequence A occurs both before and after \textit{focal} sequence B, should we place A before, after, or within \textit{focal} sequence B?
As a solution, we consider the sequences before and the sequences after \textit{focal} points as distinct, so that in this scenario, sequence A appears both before and after \textit{focal} sequence B.

\begin{figure}[!ht]
\centering
  \includegraphics[width=\columnwidth,]{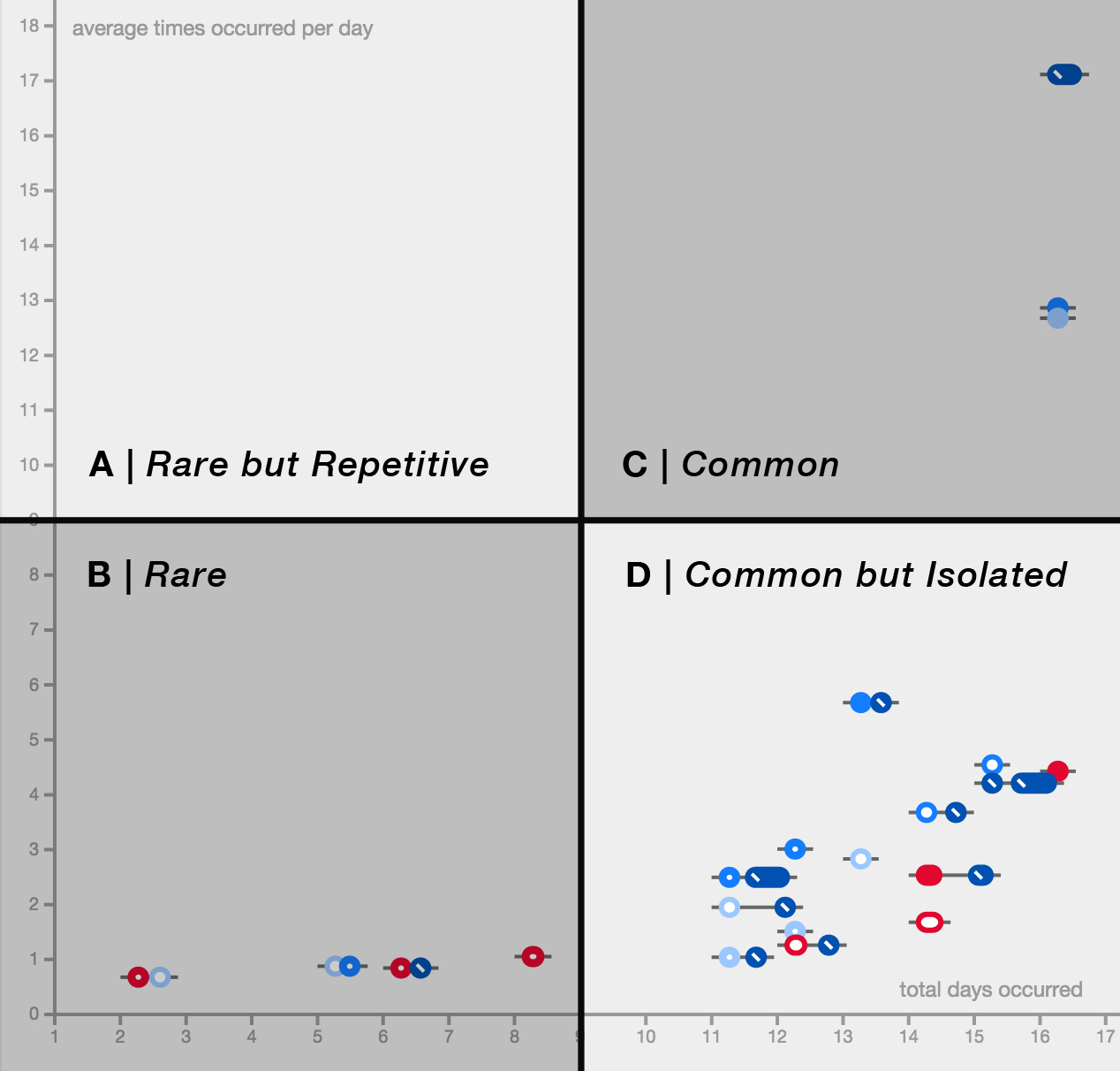}
  \caption{(Ordinary) sequences occur so frequently that they are not very interesting (e.g., walking).
  (Habitual) sequences occur on many days, but do not repeat frequently on the days that they occur (e.g., exercising once every day).
  (Rare) sequences occur very infrequently.}
  ~\label{fig:quadrants}
\end{figure}

For Chronodes' initial scatterplot, retaining inter-sequence chronology is not an issue, as sequences are positioned according to the axes.
The y-axis represents the average number of times that an event sequence occurs per day; that is, if a sequence of events tends to occur 5 times every day, its position on the y-axis is 5.
The x-axis indicates the total number of days that the event sequence is found: if the event sequence occurs for only one participant's Monday, Tuesday, and Wednesday, its position on the x-axis is 3.
As an emergent result, sequences are distributed into a two-dimensional spectrum of sequence prevalence (Figure \ref{fig:quadrants}).

\subsection{Multiple Sequence Alignments}

Unlike existing work on event-based alignment, Chronodes does not limit user interaction to single alignment on single events;
instead, it encourages the user to create multiple alignments on any number of events side-by-side. As exhibited in Figure \ref{fig:steps}, this also permits the user to find sequences of events between events of interest, unprecedented in existing event alignment techniques.

\subsection{Comparing and Cloning Cohorts}

When Jane specified the first \SEQone\:sequence from the \Stage scatterplot, this sequence appeared as a \focal sequence with its own tracks and adjacent sequences (Figure \ref{fig:steps}: Step 1).
As she continued to select more adjacent sequences like the proceeding \SEQtwo, she narrowed down on a more specific subset of the participant data: only days that included all of \SEQone$\:\cdots\:$\SEQtwo were displayed (Figure \ref{fig:steps}: Step 2)
In this way, Chronodes provides the functionality to constructively define cohorts using sequences of events.
\peter{This extends the capabilities of prior work that enables users to display participants that exhibit a specified sequence of events, but does not permit the interactive rearrangement of these events once they are set.}

\begin{figure}[!t]
\centering
  \includegraphics[width=\columnwidth,]{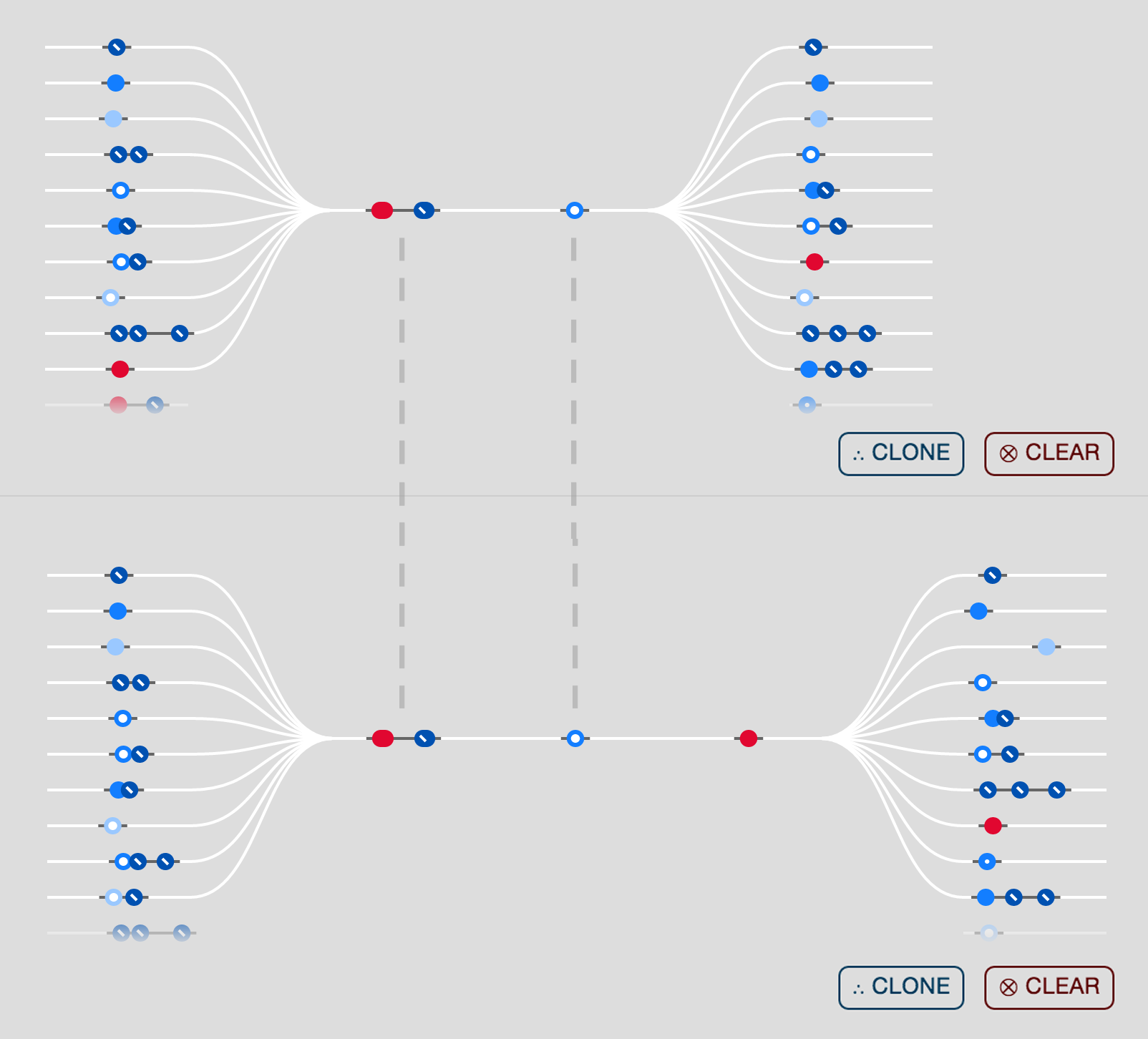}
  \caption{
  An original timeline of \textit{focal} sequences (top) that has been \textit{cloned} and then modified (bottom). By adding another \textit{focal} sequence to the bottom timeline, the event sequences before and after it are changed. In this way, Chronodes enables users to view the comparative differences between timelines, which each represents a cohort of participants.
  }~\label{fig:clones}
\end{figure}

To return to a broader subset of the participant data, Jane can either remove a \textit{focal} point, or clear the entire current timeline and return to the scatterplot.
However, she also has the option to \textit{clone} the entire timeline (Figure \ref{fig:clones}), which duplicates it on the \Stage.
By modifying the second timeline, she can compare the properties of related cohorts side-by-side. For example, by adding another smoking event (\SEQfour) to the second timeline, Chronodes displays the differences between the event-based behavior of the original cohort timeline \SEQone$\:\cdots\:$\SEQtwo and its clone \SEQone$\:\cdots\:$\SEQtwo$\:\cdots\:$\SEQfour.
As demonstrated in Figure \ref{fig:clones}, once the new \SEQfour\:is appended as a \textit{focal} sequence to the bottom timeline, the adjacent sequences before and after are updated  accordingly.






\section{Deriving Interactive Event Sequences}

Whereas Chronodes presents novel visual and interactive paradigms for analyzing chronological patterns, it is important that the system also be developed for use by mHealth researchers.
Here, we highlight Chronodes' technical contributions in providing new methods of event sequence analysis; namely, the techniques, variables, and considerations involved
in the process of deriving events, finding event sequences, and visualizing these sequences for the purposes of interaction.





\subsection{Deriving Events from mHealth Sensor Data}

\peter{Before analyzing mHealth data for chronological patterns to answer our research questions (Q1-4), each participant's physiological data streams must be represented in a consistent and comparable format.}
From each participant's array of AutoSense data streams, we extrapolated timestamped measures of (1) activity from 3-axis accelerometer chest sensors~\cite{rahman2014we}, (2) probability of stress from physiological sensors (ECG and Respiration)~\cite{hovsepian2015cstress}, and (3) instances of smoking from inertial wrist sensors (3-axis accelerometer and 3-axis gyroscope)~\cite{saleheen2015puffmarker}.
Whereas smoking episodes are described by discrete, boolean values of 1 if smoking and 0 if not, activity and probability of stress are variable between participants and need to be normalized before further interpretation.

\begin{figure}[!ht]
\centering
  \includegraphics[width=0.8\columnwidth,]{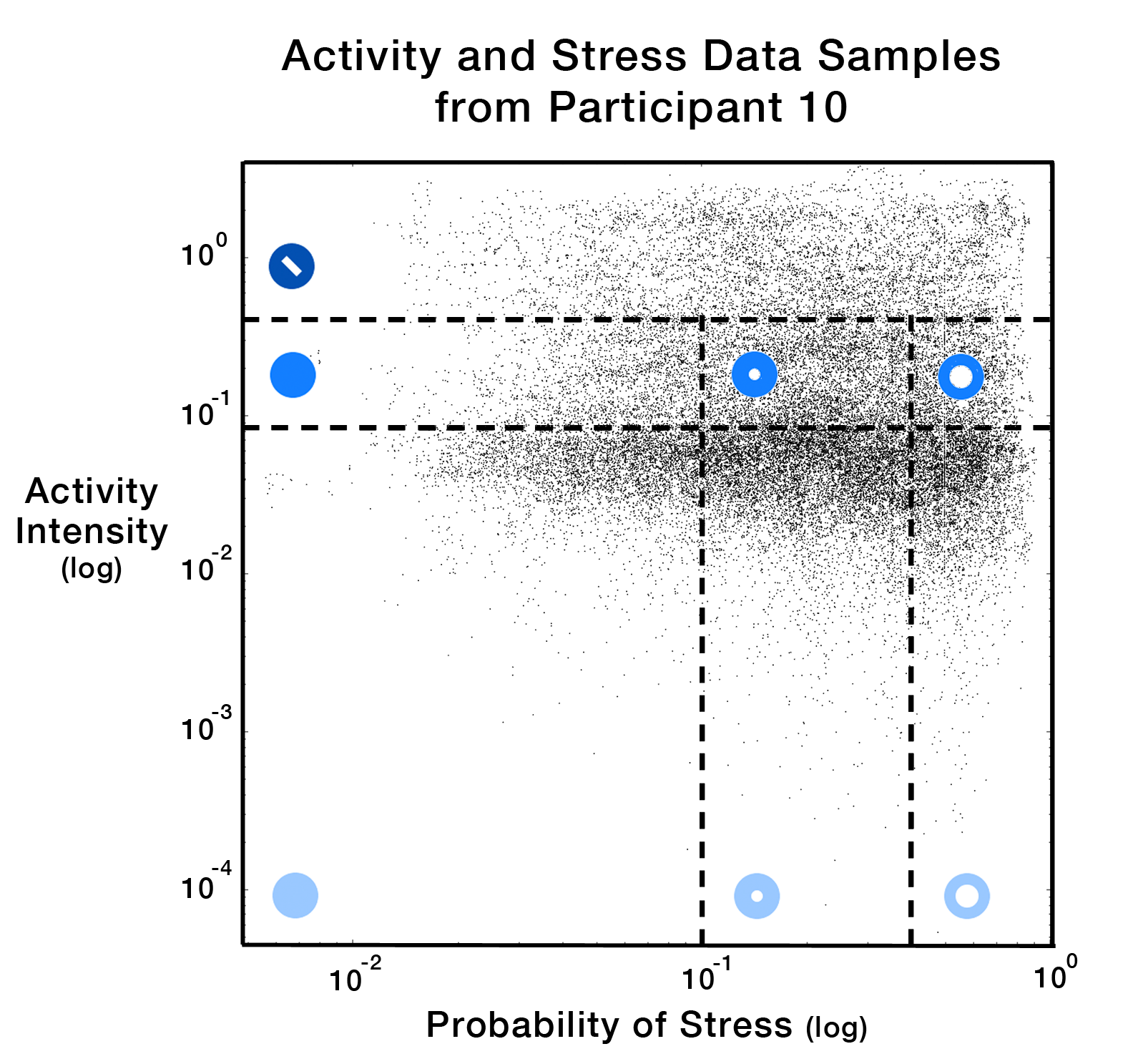}
  \caption{Participants' activity intensity and probability of stress were sampled and inferred once per second. For Participant 10, this totaled to over 168,000 data points in the scatter plot. These data points were then discretized into events with quartile thresholds (e.g., median). For events with high activity (top), stress cannot be accurately inferred.
  }
  \label{fig:plots}
\end{figure}

Once normalized, physiological data streams are classified as being either ``none,'' ``low,'' or ``high'' \peter{magnitudes at every five-minute interval of time} (Figure \ref{fig:plots}).
Our selection of this 3-phase classification model is informed by~\cite{sharmin2015visualization} which suggests that, even for experts,
a finer-grained classification increases complexity without offering additional \peter{analytical} benefits.
Whereas darkness of blue indicates activity intensity, the size of the inner white circle indicates probability of stress.
At a given point of time, probability of stress ranges from 0 to 1, or is simply -1 to indicate that data is unavailable (in the occassional event that only the activity sensor is enabled).
Also, for events where activity is high, stress cannot be accurately inferred~\cite{hovsepian2015cstress}, indicated by a white slash (\SEQsix\:in Figure \ref{fig:plots}).
\peter{Adjacent time intervals of the same magnitude} are summed into single events so that resulting events are variable in length.
As participants may not be wearing the sensors at all times, for instance during sleep, consecutive events that are separated by long gaps of no data collection at all are not merged \peter{past a user-defined threshold (60 minutes of sensor inactivity by default)}.

Before mining participants' daily routines for \peter{recurrent} sequential patterns, we \peter{should} define what constitutes a `day:' although the majority of participant activity in our study occurred during daytime hours, some instances of irregular sleeping patterns and nighttime activity were present.
Some participants even continued to wear the sensors during sleep.
This justifies calculations that determine participant waking hours, but presently, for consistency and simplicity, Chronodes defines a `day' as a hard 24 hour time period from midnight to midnight,
necessarily accounting for behaviors that repeat at consistent times across days.

\subsection{Deriving Frequent Sequences From Events}
\peter{Once participant data streams are represented as a continuous series of events, sequence mining provides Chronodes users with meaningful chronological relationships and sequences of events that can be manipulated interactively.}
Upon launch, Chronodes finds daily event sequences present throughout all participant events using the PrefixSpan~\cite{Pei:2001:prefixspan} sequence mining algorithm.
PrefixSpan retrieves patterns that occur frequently amongst sequential events, and is an efficient algorithm for mining frequent sequences
from a large number of individual event strings (i.e., a large number of 24 hour days). Algorithms of this kind are increasingly useful as we scale up to more participants.
Before the mined frequent sequences can be rendered as visual components that the user can interact with, we must first modify PrefixSpan to consider variations in \textit{repetitive}, \textit{gapped}, and \textit{closed} sequence mining.

When PrefixSpan runs, it returns a list of frequent sequences ranked by frequency, with their positions in the participant data. Importantly, the algorithm does not mine for \textit{repetitive} frequent sequences: if a smoking episode occurs three times every day, PrefixSpan only recovers the first occurence per day. To extend the algorithm to find \textit{repetitive} sequences, we use the locations of the initially found frequent sequences to search the remainder of the 24 hour day for repetitions.

Also by default, PrefixSpan mines for \textit{gapped} event sequences: events can be considered as a part of the same sequence even if they are separated by many other events (a `gap'). This is not ideal for scenarios where healthcare analysts are only interested in the events immediately preceeding or proceeding smoking relapse; however, preventing gaps entirely would not allow our users to understand relationships between event behaviors that occur hours apart. To suit this range of use cases, we set a maximum gap parameter that Chronodes users can adjust concurrently.

Ultimately PrefixSpan does not limit mining to \textit{closed} sequences; for example, it redundantly returns both the \textit{closed} \SEQfive\:, and its subset \SEQone.
Although we can modify PrefixSpan to return only the \textit{closed} frequent sequences, Chronodes' \peter{design} lends interactive capabilities that allow us to do something more comprehensive.
\peter{By initially providing} only the shorter frequent sequences that have a longer variation (in our example, present only \SEQone), by selecting this sequence as a \textit{focal} sequence we can see all of the variations that are related to it. As Chronodes displays the sequences that come after the \textit{focal} sequence once it is selected (like \SEQseven), it is also displaying longer variations of the \textit{focal} sequence as a side-effect (\SEQone$\:\cdots\:$\SEQseven\:is equivalent to \SEQfive). \peter{In doing so, we enable a new interactive method for exploring event sequences mined with frequent sequence mining, whereby users can constructively narrow in on related sequences. Additionally, as an emergent result, we can keep the frequent sequence `entry points' on the scatterplot simple.}

\subsection{Interacting with Frequent Sequences}

Chronodes maps every mined frequent sequence to its recurring location in the event data, so that mousing over or selecting each sequence indicates its distribution in the \Orchestra (Figure \ref{fig:fullscreen}). Also, as every frequent sequence is related to a particular cohort of participants (either all or some participants have the sequence), we can use information about the associated cohort to limit our searches for adjacent sequences. Any participant or day outside of the currently specified cohort does not need to be searched for sequences, thereby limiting use of computational resources as we scale to more participants and data streams.

Specific care is taken to ensure that, when a frequent sequence is selected in Chronodes, the user can understand the effects of their action on the visualization. Upon selection, sequences animate to the center as the nearby sequences fade away, and the new \textit{tracks} grow outward.
After these animations complete, the \Orchestra updates to indicate the \textit{focal} sequences and new cohort of participants being observed.

\subsection{System Implementation}
Chronodes's front-end visualization component is web-based and written in JavaScript (jQuery\footnote{jquery.org}, React\footnote{facebook.github.io/react}, D3\footnote{d3js.org}), and served by a Python web server (web.py\footnote{webpy.org}).
The front-end interface sends API requests to the web server that returns data processed for display.
Timestamped inferences about participant activity, stress, and smoking episodes are stored in an SQLite database\footnote{sqlite.org}, for its cross-platform compatibility, integration with Python, and support for the dataset size of our pilot study.


\section{User Study with Health Experts}

We conducted an informal pilot investigation with 20 behavioral, biomedical, and computational health experts coming from a large research team to gain insights into the efficacy and limitations of Chronodes.
Through the study 
we intended to understand how Chronodes may help them with mHealth data exploration, pattern discovery, and decision making. The interdisciplinary nature of the research team presents the unique opportunity for us to gain insight into how Chronodes may be used by experts with diverse backgrounds. 
Table \ref{tab:simple} highlights some of their domain expertise and experience.
We refer to these participants as ``experts'' to avoid confusion with the AutoSense participants (from whom the mHealth data was collected).



\begin{table}[t]
\small
\sffamily

\centering

\begin{tabular}{l l r}
\toprule

\textbf{ID} & \textbf{Domain Expertise} & \textbf{Years in Field}\\

\midrule

A & Health psychologist 
& 13 \\ 

B & Statistician (focus: clinical design, analytics) 
& 26\\ 

C & Behavioral health researcher 
& 38 \\ 

D & Machine learning 
& 10 \\ 

E & Health informatics 
& 31 \\ 

F & HCI, human-centered computing 
& 26\\ 

G & Sensor and hardware designer 
& 16 \\ 

H & Clinician \& health informatics 
&  25 \\ 

J & mHealth software designer 
& 13 \\ 

K & mHealth software architect 
& 12 \\

\bottomrule
\end{tabular}
\caption{Expert participants,
their diverse domain expertise and years in fields (12--38 years).
These experts' characteristics are representative of the 20 experts in our study.
}
\label{tab:simple}
\end{table}


\subsection{Method}

Each study session began with a demonstration of the Chronodes interface.
The participants were welcome to ask questions at any time.
After the demonstration, the participants were instructed to think aloud in describing their perspectives and criticisms of the interface's features.
Comments were recorded and organized by which interface feature they pertained to: (1) event derivation and representation, (2) frequent sequence derivation, representation, and placement, (3) mining and placement of adjacent sequences around \textit{focal} sequences, and (4) the use of multiple \textit{focal} sequences.
Due to the nature of our informal study, we did not record audio of the demonstrations.
We took notes to capture feedback on the usability, effectiveness, as well as limitations and possible improvements of the system. 
Remarks included in this section are paraphrased summaries of this feedback.

\section{Results and Implications for Future Research}
In this section, we summarize insights gleaned from the user study, present open challenges, and provide recommendations for future visualization and analysis research with mHealth data.

\subsection{Representing mHealth Data}

The challenge of analyzing and representing mHealth data is one of variability (continuous quantitative data streams), uncertainty (missing or incorrect measurements), high volume (data from many sensors simultaneously), and high dimensionality (many participants and many data streams) at once.
To represent continuous mHealth data streams across many participants effectively, focused investigations into each of these factors are necessitated for future research.


\subsubsection{Leveraging Temporal Variation during Event Derivation}

Drawing on prior works \cite{Monroe:2013:temporal,Malik:2015:cohort},
Chronodes represents mHealth data as discrete events,
\peter{which supports the visualization and analysis of chronological patterns, their frequencies, and their relationships.}
In \peter{EHR} scenarios where health data is already described by discrete events (e.g., medications administered in asthma treatment in EventFlow~\cite{Monroe:2013:temporal}, hospital administration events in CoCo~\cite{Malik:2015:cohort}), \peter{discretization from continuous data streams to discrete events is not required.
However, working with quantitative data streams in mHealth provides the opportunity to define more complex event types, leveraging temporal variation to define discrete events that capture the ways physiology changes over time.}

For example,
Expert H described that when an event occurs, health analysts are often interested in the residual effects of the event over time, and how these effects might in turn affect our understanding of the events that follow.
Additionally, some events are better described as fluctuating, quantitative values altogether.
In other words, being able to investigate the \textit{underlying causes} or \textit{emergent effects} of an mHealth event is useful for understanding why it is present in the first place, and what kinds of implications it has.
For this,
Expert H provided the example of stress: although we may certainly represent a high-stress episode as happening at a discrete point in time, it may be more effectively represented as a state with a variable magnitude, and with variable effects that are sustained until after the high stress is sensed.

\begin{figure}[!ht]
\centering
  \includegraphics[width=\columnwidth,]{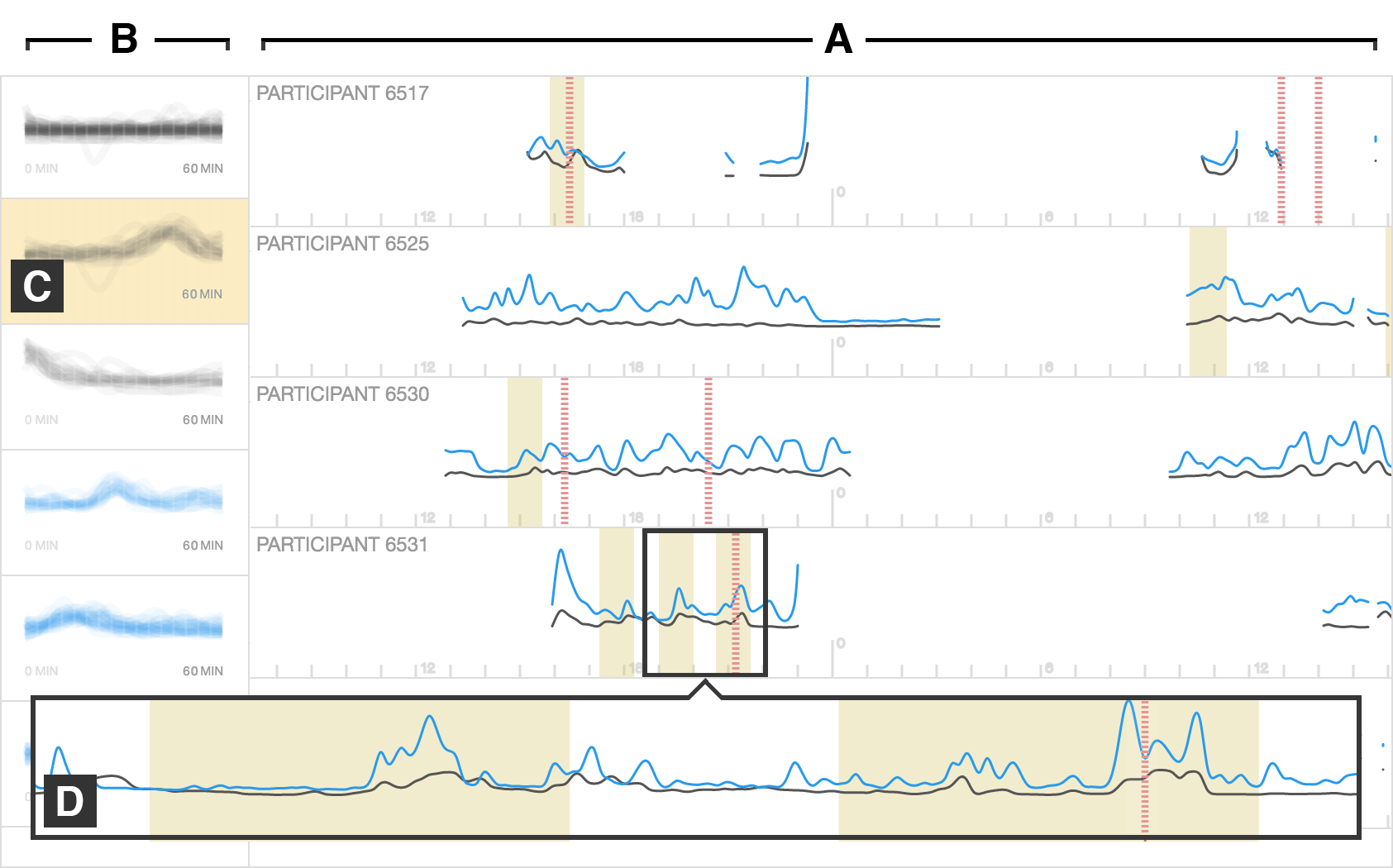}
  \caption{
  (A) The event derivation interface showing probability of stress (gray), activity intensity (blue), and smoking episodes (red vertical dotted lines) for multiple participants. 
  (B) The sidebar displays common temporal trends (motifs) identified from this data using time series motif mining. 
  (C) The user selects a motif to highlight where it occurs in the data. 
  (D) Zoomed-in view of a three-hour region showing two instances of the highlighted motif (gray border added to enhance clarity).
  Motifs selected from the sidebar can be defined as events for use in Chronodes.
  }~\label{fig:motifs}
\end{figure}

Describing mHealth data as temporal features \cite{Gschwandtner:2011:carecruiser} or motifs \cite{Chiu:2003:probabilistic} are potential routes for addressing this challenge.
\peter{To this end, we enhanced Chronodes with a event derivation system (Figure \ref{fig:motifs}) to discover temporal patterns in mHealth data, which can then be organized in Chronodes.
To find these patterns, the event derivation system employs the Symbolic Aggregate approXimation and Vector Space Model (SAX-VSM)~\cite{senin2013sax} to represent the time series in the mHealth data (e.g., stress levels, activity intensities) as temporal segments,
that are then clustered into groups with k-means~\cite{goutte1999clustering,phu2011motif}.
After discovering these groups of trends, or \textit{motifs}, across participant records, the event derivation system allows users to examine where these motifs occur, and select any number of them.
Once selected, motifs can be defined as events for use in Chronodes, represented as glyphs of the sidebar motifs.}

\subsubsection{Addressing Uncertainty in mHealth Data}

mHealth data presents a unique relationship between the high-resolution data points recorded and the high-level human behaviors that can be inferred from them at varying degrees of certainty.
In this way, Expert H was interested in how Chronodes might portray the relationships between ground truth sensor events and the behavioral variables that can be inferred from them (e.g., high stress inferred from increased heart rate).
Similarly, prior research suggests that there is a discrepancy between a self-reported event-timing and the actual event-timing: smokers often report smoking episodes prior to or after the actual smoking event~\cite{saleheen2015puffmarker}.
Designing a temporal visualization that \peter{mediates conflicts between} self-reported events and sensor-collected objective event points thus introduces novel challenges that need to be further investigated.
Recent work has made a first attempt \cite{sharmin2015visualization}, but further research is warranted to find optimal solutions as to how to best integrate self-reported data into sensor recorded data, \peter{and to visualize these integrations}.

\subsubsection{Sense-making from High Volume mHealth Data}
Experts C and J commended Chronodes' approach to making sense of high-volume event data by way of interaction with event sequences, which was echoed by other experts consistently.
These remarks suggested Chronodes' niche role in exploring mHealth data interactively with event sequences, particularly as these methods relate to other existing healthcare analysis techniques that focus on event-based analysis (e.g., \cite{Monroe:2013:temporal,Malik:2015:cohort}).
Ultimately we can use the existing interactive techniques in Chronodes to mine high-volume mHealth data more effectively,
learning from how tools like Progressive Insights~\cite{stolper2014progressive} permit exploration of data as it is being mined.

\subsection{Context and Homologous Events}

\subsubsection{Instantial Context} 
Expert D noted that in some cases events may be identical in terms of their content, but different in the context of how they were registered or defined.
A specific example given to us was: how do we denote the difference between a self-reported smoking event, and one detected by carbon monoxide sensors?
Of course, as these events are instantiated in different ways we could represent them as visually distinct, but in developing Chronodes we would prefer to design a more robust way for users to understand the relationships between, and existence of, these homologous events.

\subsubsection{Situational Context}
Expert D described the differences between events that might depend on location and spatial context, like between a smoking event at home and at work.
Already we are designing a next iteration of Chronodes to account for these situational contexts.
As our frequent sequence mining algorithm supports events that occur simultaneously, we are able to consider and represent `smoking' event (behavioral) occurring `at home' (spatial) at once.
As a result, these events in parallel provide context to one another by being present at the same time.
However, visualizing multiple contextual events in parallel may confound which events relate to one another in the first place, especially as we scale to more event types.
Further research is needed to determine the kinds of computational, visual, and interactive techniques required to indicate how a variety of mHealth events relate to their situational contexts.

\subsubsection{Temporal Context}
Expert A is working to determine false positives in sensor-detected smoking episodes, and indicated a need to understand the differences between events that depend on time, like between a smoking event in the morning and at night.
We are also designing a solution for this kind of temporal context:
as the \Orchestra represents a timeline, we can use it as an interactive filter to specify the time regions that we are interested in.
In our scenario, Jane should be able to highlight the morning in the \Orchestra, and choose a smoking episode \textit{focal} sequence limited to that specified time area.
In this way, she would see only the events that occur before and after smoking episodes in the morning.
This is a principal update that we are working to develop into our next version of Chronodes.

\subsection{Extensibility to Other mHealth Analysis Scenarios}

\peter{Across the study sessions} Chronodes was deemed generalizable to a variety of uses in healthcare and mHealth data analysis. \peter{We describe these applications in this section to motivate future work in this area.}

\subsubsection{Behavioral Analysis}
Expert B commented on the applicability of the Chronodes interface to behavioral scientists, and suggested to perform evaluative studies with this research area next.
Particularly, after we had explained how timelines represented cohorts and could be cloned, Expert B described the value of imposing constraints on the cohort timelines other than event order.
A specific example given was investigating the differences in event chronology between male and female participants.
This is a comparative strategy emphasized in prior motivating works like CoCo~\cite{Malik:2015:cohort}, and when paired with \peter{online} event sequence specification is foreseeably powerful.

\subsubsection{mHealth Interventions}
Expert B also expressed excitement for Chronodes' potential applications in investigating the effects of just-in-time adaptive interventions.
Specifically, marking different kinds of interventions as events and specifying them as \textit{focal} sequences would be a significant step to not only planning interventions, but also understanding their outcomes across participant groups.
For other end users, Expert J suggested that Chronodes could play an important role in clinicians interacting with these participants, where participants could be asked to recall why they tended to exhibit certain behaviors after a specified event.

\subsubsection{Health Sensor Development and Validation}
Expert F described Chronodes as a useful tool for mHealth sensor developers that need to understand the relationships between the data elements that they record.
Expert C, likewise, described how Chronodes could be used to help sensor developers identify false positives from sensor alerts and corresponding user reports.
With this insight we are deeply considering Chronodes' applications not only in end-user behavioral analysis, but also in the validation of mHealth technologies in the first place.


\section{Related Work}
Our research builds on prior works in multiple disciplines, from visualization, health research, and data mining.

\subsection{Visualization of Multiple Timelines}

Aggregating the event data of multiple timelines into a single timeline can be difficult to make sense of: the order of events from one timeline to another is completely uncorrelated.
This challenge is customary in healthcare analysis: apart from needing to understand the relationships between a patient's various physiological records, analysts must do the same between thousands of patients on end \cite{bar2004biostream}.
Fundamentally, visualizing dissimilar records side-by-side 
\moushumi{proposed} in \cite{fouse2011chronoviz} enables users to 
\moushumi{identify} high-level correlations between the distinct data streams. However, this approach becomes increasingly intractable as we scale to more patients and sensor kinds. In an effort to unify many patient timelines, CareFlow~\cite{perer2013data} visualizes the outcomes of 50,000 patients in a single, tree-like timeline. Although this approach provides an expressive high-level overview of similar patient groups, it has limited support for \peter{the specification of behavioral groups and repetitive event sequences}.
Whereas Chronodes provides a summary of participant data in this fashion (Figure \ref{fig:figure1}A), its concentration is a new sequence-based overview that reveals repeating event patterns at a glance.







\subsection{Event Alignment}

Tools like EventFlow \cite{Monroe:2013:temporal}, LifeFlow \cite{Wongsuphasawat:2011:lifeflow}, as well as others in non-health related fields 
\moushumi{such as} Experiscope \cite{guimbretiere2007experiscope}, allow users to select a specific event (e.g., drug A prescribed), thereby displaying the events that occur both before and after the selected event, for all data records.
This technique of `event alignment' is useful for extrapolating event causation and chronological trends, and also for organizing event flows around a consistent visual reference point.

Still, even with alignment, the abundance of data presented by these interfaces, especially for large datasets, is complex and difficult to understand at a glance.
To mitigate this visual overload, existing tools rely on techniques of simplification, such as Find-and-Replace \cite{Monroe:2013:temporal}, which can compress multiple user-specified events into one.
Although this kind of simplification reduces the number of visual elements on the screen, it hides information as the user is repeatedly working toward an increasingly simplistic representation---it is difficult to recover how the simplified data stream represents the original data~\cite{Monroe:2013:temporal}.
\peter{With an interactive strategy similar to event alignment, Chronodes reveals event-based patterns from aggregated data streams without the visual complexity of presenting mHealth data records as-is.}
\peter{Then, as opposed to iteratively consolidating events to simplify the visualization of all participant records, Chronodes leverages the properties of frequent event sequences to depict patterns of behaviors that can be constructed and explored interactively.}

\subsection{Cohort Analysis}

An alternate approach to understanding the trends of aggregated mHealth data streams is to consider groups of patients as cohorts, or individuals that share certain properties \cite{basole2015understanding,Malik:2015:cohort}.
\hide{Through  discussions with doctors and researchers on our team, we found that these cohorts were not only difficult to define, but also to represent and compare intuitively.
Additionally, in these discussions we acknowledged that defining cohorts with universal factors (e.g., age, gender) was insufficient for revealing differences in event-based behavior between cohorts.
How might we define an individual who lapses in the morning as belonging to a different cohort than another who lapses at night?} \hide{probably remove this, a reviewer had a problem with our claim that cohorts are difficult to define, and Chronodes does not \textit{necessarily} provide a solution to the question posed}
In an effort to comparatively assess sequences of events between cohorts, projects such as CoCo (Cohort Comparison)~\cite{Malik:2015:cohort} show the chronology of events associated with two user-defined cohorts.
Whereas this effectively reveals the different properties and event sequences between specified cohorts, CoCo is limited to specifying and comparing two cohorts \peter{(e.g., male and female)}, and does not enable the user to redefine cohorts groups by arranging sequences of events.
Chronodes builds on the cohort comparison methodology
by allowing the user to define cohorts constructively: \peter{rather than specifying cohorts for comparison and viewing their sequences of events, Chronodes allows cohorts to be specified by an arrangement of the events themselves.}
By pairing this functionality with event alignment, Chronodes enables the user to fluidly explore cohorts and their associated chains of events in tandem.

\subsection{Sequence Mining and Pattern Matching}

Building machine learning and pattern matching algorithms into interactive visualizations presents promising opportunities for enhancing human pattern-finding abilities. \cite{hochheiser2002dynamic} and \cite{buono2005interactive} demonstrated how users could select specific patterns in quantitative data streams, and see where else they were found. For event-based data, \cite{fails2006visual} developed a query system to find patterns across multiple event histories.
\cite{Gschwandtner:2011:carecruiser} provides users with more options to specify how these patterns are defined, and lends detailed representations of how patterns compare.

Acknowledging the potential of frequent sequence mining in dense event-based health data, ActiviTree~\cite{vrotsou2009activitree}
renders frequent event patterns found throughout participant data as a visual tree of `common' event sequences.
Frequence~\cite{perer2014frequence} bridges this technique with its precursor CareFlow's~\cite{perer2013data} representation of events and connecting edges, and then enables specification of cohorts by selecting sequential branches.
Allowing users to extend events sequences as they are retrieved, Progressive Insights~\cite{stolper2014progressive} offers an alternate, constructive methodology to specifying event sequences.
Whereas TimeStitch~\cite{polacktimestitch} designed new interactive paradigms for specifying and manipulating event sequences in a constructive, non-linear fashion, \peter{Coquito~\cite{krause2016supporting} enabled the same constructive definition of behavioral groups for cohorts.
Chronodes builds on this foundation of work with a constructive event-based querying system, enabling the exploration and discovery of frequent patterns across multiple continuous timelines of mHealth records.}


\section{Conclusion}

We presented Chronodes, a system that unifies data mining and human-centric visualization techniques to support the explorative analysis of longitudinal mHealth data.
Chronodes extracts and visualizes frequent behavioral sequences and promotes them as interactive elements, enabling health researchers  to interactively define, explore, and compare groups of participant behaviors using event sequence combinations.
We evaluated Chronodes with an informal study with 20 expert health researchers, and determined the application's utility across a variety of healthcare and computational disciplines.
From these insights, we are continuing to develop Chronodes for applications to other health data analysis scenarios, including understanding the health events surrounding congestive heart failure and diabetes management, and observing the effects of just-in-time adaptive interventions from perspectives in behavioral science.

\bibliographystyle{SIGCHI-Reference-Format.bst}
\bibliography{citation}

\end{document}